# Comment on **Universal Reduced Potential function for Diatomic Systems**

Since QM fails on a low-parameter universal function (UF) for potential energy curves (PECs), searches for the UF remain important [1-7]. Despite its title, Xie and Hsu's claim [1] is invalid. Their 200 bonds between atoms with closed/S-type shells cover H, columns 1-2 and 8 with noble gases. Since column 7 with univalent atoms, covering 50% of *common* bonds, is excluded, while 112 (56 %) *uncommon* bonds with noble gases are included [1,2], claim [1] is invalid by *reductio ad absurdum*. If it were valid, its universal bond is typified with *bound inert gases*, which are *non-bonding*. This self-contradictory result falsifies [1]. Further remarks are pertinent.

Ref. [1] uses (i) Dunham's potential, scaled by $D_e$, giving covalent Sutherland parameter [3-7]

$$\Delta=\Delta_{cov}=\tfrac{1}{2}k_e r_0^2/D_e \qquad (1a)$$

and (ii) Varshni's procedure with F and G [7]. For the UF, smooth G(F) as well as F($\Delta$) and G($\Delta$) plots should encompass observed constants [3-7]. However, for 300 bonds with H and atoms in all columns 1-8, F($\Delta$) and G($\Delta$) are scattered; only G(F) is relatively smooth [4]. While [3-7] all find scattering for many bonds in [1], Xie and Hsu give smooth plots in their Fig. 2 [1]. To understand this contradiction, Fig. 1 here expands on G($\Delta$) for G≤100 [1] with data [5] added. The scattering in Fig. 1, reported in [3-7], is obscured in [1] by data-compression to give *seemingly smooth plots* in the *small graphs*

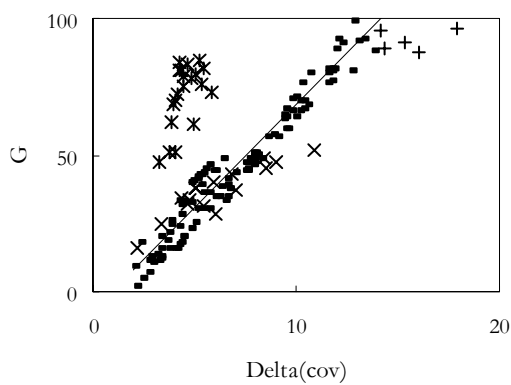

Fig. 1 G($\Delta$) for G≤100: [1] (-); [5] (+) halogens $X_2$, (*) salts MX, (x) bonds with H,M,X

of their Fig. 2. This procedure is questionable, to say the least.

The full line in Fig. 1 reveals that 200-bond fit G=7,2422$\Delta$ [1] is rather poor. In fact, it hardly differs from analytical result G=7,3333$\Delta$ for Rydberg's *not universal* function [7].

In [1], the large annoying gaps for halogen bonds [5] (see Fig. 1) are avoided by *bond selection* but this leads to a self-contradictory universal bond (see above). Later, but not in [1], Xie with Heaven concedes [8] that ionic MX *deviate* markedly from universality-rule [1] but not that this voids claim [1].

Ref. [1] argues that, *contrary to* [4], (1a) is a valid scaling aid. This 300-bond study [4] reveals that *ionic* Sutherland parameter

$$\Delta_{ion} = \tfrac{1}{2} k_e r_0^2 / D_{ion} \qquad (1b)$$

with ionic energy $D_{ion} = e^2/r_0$ [5] is *superior* to (1a). In fact, for *all bonds with all univalent atoms in the Table*, $\Delta_{ion}$ makes both $G(\Delta_{ion})$ and $F(\Delta_{ion})$ as smooth as $G(F)$, which probes the UF for *common* bonds [4,5]. If large gaps in Fig. 1 persist, something must be wrong with $\Delta_{cov}$ (1a) and/or Dunham theory [5]. With $\Delta_{ion}$ not mentioned and [5] omitted in [1,2], this scaling controversy in [1] is futile.

Since $\Delta_{ion}$ unifies *common* bonds, their unifying *ionic* function should at least be accurate for *prototypical covalent* $H_2$, also in [1]. This calls for an accuracy test for $H_2$ using observed $r_0=0.7414$ Å and $\omega_e=4402$ cm$^{-1}$ [9]. While the *empirical* fit of the *3-parameter* potential in [1] gives $\omega_e=4656$ cm$^{-1}$, *parameter-free* ionic Kratzer potential $U_K(r) = \tfrac{1}{2}(e^2/r_0)(1-r_0/r)^2 = \tfrac{1}{2} k_e r_e^2 (1-r_0/r)^2$ [4,6,10] gives *analytically* $\tfrac{1}{2} e^2/r_0 = 78500$ cm$^{-1}$, $k_e = e^2/r_0^3 = 5.7 \cdot 10^5$ dyne/cm and $\omega_e = 4390$ cm$^{-1}$!

With bonds between all univalent atoms unified by an ionic Kratzer-type UF, their PECs favor ionic bonding, implying that *prototypical covalent structure* $H_2$ acts like *a pair of ionic structures* [H$^+$H$^-$; H$^-$H$^+$] [4]. Despite appearances, this *old* model is not self-contradictory and has *new* implications [4,6,11], which false claim [1] can never have.

G. Van Hooydonk, Ghent University, Belgium

I thank J.F. Ogilvie, P. Hajigeorgiou, H.G.M. Edwards and others for support.